\documentclass[prb,twocolumn]{revtex4}
\usepackage{array,amsmath,amssymb}

\begin{document}

\title{Short-range correlations and spin-mode velocities in
ultrathin one-dimensional conductors}

\author{Michael M. Fogler}


\affiliation{Department of Physics, University of
California San Diego, La Jolla, California 92093}

\date{\today}

\begin{abstract}

In ultrathin wires positioned on high-$\kappa$ dielectric substrates or
nearby metallic gates, electrons can form strongly correlated
one-dimensional fluids already at rather high electron densities.
The density-density correlation function, charge compressibility, spin
susceptibility, and electron specific heat of such fluids are calculated
analytically. The results are relevant for transport and thermodynamics
of carbon nanotube field-effect transistors and semiconductor
quantum wires.

\end{abstract}

\maketitle

There is a long-standing theoretical prediction that one-dimensional
(1D) electrons do not obey the conventional Fermi-liquid theory but
instead form a Luttinger liquid (LL)~\cite{Haldane_81} whose fundamental
degrees of freedom are bosonic modes that separately carry electrons'
charge and spin. Experimental verification of the LL theory has proved
to be a challenge. There is, however, evidence for a LL-specific
suppression of tunneling into carbon nanotubes (CN)~\cite{Bockrath_99}
and a spin-charge separation in quasi-1D organics.~\cite{Claessen_02}
Vexing questions remain: (1) Is there a way to demonstrate a truly
dramatic departure from the Fermi-liquid in real 1D systems? (Aside from
the special case of quantum Hall edge states.~\cite{Chang_03}) And if
so: (2) What theory should one use to describe such a regime in the
presence of long-range Coulomb interactions? Finally, (3) What phenomena
can one expect from this deeply non-Fermi-liquid state? Below we propose
that the desired strong coupling regime can be realized if an {\it
ultrathin\/} wire, e.g., a CN is (i) placed on a high-$\kappa$
dielectric substrate or (ii) brought close to a metallic gate. In
contrast, the traditional prescription for obtaining a strongly
correlated regime is to lower the electron density $n$ thereby
increasing the Coulomb coupling constant $r_s = 1 / 2 n a_B$, where $a_B
= \hbar^2 \kappa / m_\ast e^2$ is the effective Bohr radius.
Unfortunately, this route quickly runs into the problem of localization
by disorder, e.g., random charges on the substrate. The advantage of our
proposal is that the Coulomb potential of these charges would be
strongly screened, whereas interactions among electrons would be
affected much less, as shown below.

We begin with the dielectric substrate case. We call a wire of radius
$R$ {\it ultrathin\/} if $R \ll a_B$. Experimentally, enormous $a_B / R
\sim 10^4$ ratios are achievable in devices that use zigzag
CNs~\cite{Saito_book} placed on the SrTiO$_3$
substrate~\cite{Javey_02} ($R\sim 1\,\text{nm}$, $\hbar^2 / m_\ast e^2
\sim 50\,\text{nm}$, $\kappa \sim 200$). We show that in such an
ultrathin wire Coulomb correlations are enhanced by a large parameter
${\cal L} = \ln (a_B / R)$ and a correlated regime --- Coulomb
Tonks Gas (CTG) --- appears in a window ${\cal L}^{-1} \ll r_s \ll 1$ of
\textit{low\/} $r_s$. The CTG can be defined as the state where on all
but exponentially large lengthscales, $x < x_* = a_B \exp(\pi^2 / 2
r_s)$, electrons behave as a gas of impenetrable but otherwise free
fermions. At such $x$ the LL theory, being an asymptotic {\it
long-wavelength\/} theory, has no predictive power, and the alternative
method presented below is needed; $x_\ast \sim 1\,\mu\,{\rm m}$ should
be achievable in current CN devices.~\cite{Javey_02} The CTG owes its
name to a certain similarity it enjoys with the Tonks-Girardeau gas of
1D cold atoms.~\cite{Paredes_04}

We find that to the leading order in $1 / {\cal L} \ll 1$, the
short-range {\it density\/} correlations in the CTG are identical to
those of a one-component free Fermi gas. The {\it spin\/} correlations
are the same as in the 1D antiferromagnet. We show that the CTG
possesses a number of properties akin to the $r_s \gg 1$ 1D Wigner
crystal,~\cite{Schulz_93} including a negative compressibility, a high
spin susceptibility and electron specific heat, and also anomalous
finite-temperature transport~\cite{Matveev_04} and
tunneling~\cite{Cheianov_04} properties. Thus, although the 1D Fermi
gas, the CTG, and the Wigner crystal are not different thermodynamical
phases (they all are LLs), significant quantitative differences in
short-range properties of the electron system in these three regimes
cause sharp crossover changes in the observables, similar to the boson
case.~\cite{Paredes_04}

Let us proceed to the derivation of these results. The crucial insight
comes from the two-body problem. Consider two electrons with a relative
momentum $q \sim 1 / a = 1 / 2 r_s a_B$ interacting via a model
potential $U(x) = e^2 / \kappa (|x| + R)$. We face the following puzzle.
If $r_s \ll 1$, we have the inequality $e^2 / \kappa \hbar v \ll 1$, so
that the kinetic energy greatly exceeds the characteristic Coulomb
energy ($v = \hbar q / m_\ast$ is the relative velocity). Naively, one
may expect that the Coulomb potential should be a small perturbation.
But the reflection coefficient computed in the first Born approximation
is equal to $i \ln (R q) / a_B q$, which is large in the range of $r_s$
that corresponds to the CTG. To resolve this puzzle one has to separate
the effects of the tails of the Coulomb potential (large-$x$) and of its
sharp increase at the origin (small-$x$). Indeed, let us examine the
Schr\"odinger equation for the wavefunction $\phi$ of the relative
motion
\begin{equation}
-\phi^{\prime\prime}(\xi) + \frac{r}{|\xi| + \alpha} \phi
-\frac14 \phi = 0,
\label{phi_eq}
\end{equation}
where $\xi = q x$ is the dimensionless separation, $r = 1 / q a_B$, and
$\alpha = R q$. We focus on the case $r \sim r_s \ll 1$, $\alpha \ll 1$.

The general solution of Eq.~(\ref{phi_eq}) is given by
\begin{equation}
\phi(\xi) = A_s W_{i r, \frac12}(i |\xi| + i \alpha)
          + B_s W_{-i r, \frac12}(-i |\xi| - i \alpha),
\label{phi}
\end{equation}
where $s = \text{sgn}(\xi)$ and $W_{\mu, \nu}(z)$ is the Whittaker
function.~\cite{Gradshteyn_Ryzhik} The constants $A_s$ and $B_s$ must be
chosen to ensure the continuity of $\phi$ and $\phi^\prime$ at $\xi =
0$. The Whittaker function has the asymptotic
behavior~\cite{Gradshteyn_Ryzhik}
\begin{equation}
W_{i r, 1 / 2}(i \xi) \sim \exp[-i (\xi / 2) + i r \ln(\xi) - \pi r],
\quad \xi \gg r.
\label{W_asym}
\end{equation}
Due to the logarithm in Eq.~(\ref{W_asym}), which results from the
slowly decaying $1 / |x|$-tail of the interaction potential, the
scattered states are not exactly plane waves. Thus, if $t(q)$ denotes
the transmission amplitude, its phase depends on the distance $x$ from
the origin to the points where the scattered states are measured, $t(q)
\propto \exp[-i r \ln (x q)]$. However, in a wide range of $x$, from the
classical turning point $x \sim r / q \sim a^2 / a_B$ to an
exponentially large distance $x \sim q^{-1} \exp(1 / r) \sim a \exp(1 /
r_s)$, this dependence of $t(q)$ is very slow. The phase shift
accumulated over this entire interval of $x$ is small and can be
ignored. This is consistent with Coulomb potential being a small
perturbation at such $x$. Note that the point $x \sim 1 / q$, which
corresponds to the the characteristic interelectron distance $a$ in the
many-body problem, is safely within the indicated range of $x$.
Therefore, for our purposes we can define the transmission coefficient
by $t(q) \equiv B_+ / A_-$ at $A_+ = 0$. With this definition, one can
show that $t(q)$ is given by
\begin{equation}
t(q) = i \exp(-\pi r)
[ ({d} / {d \alpha}) W_{-i r, 1 / 2}^2(-i \alpha)]^{-1},
\label{t_general}
\end{equation}
which entails (cf.~Ref.~\onlinecite{Gradshteyn_Ryzhik})
\begin{equation}
t(q) = \frac{i q}{i q - c(q)},\quad c = -\frac{2\ln (R q)}{a_B},
\quad q a_B \gg 1.
\label{t_small_r}
\end{equation}
Thus, there exists a window of momenta, $1 / a_B \ll q \ll {\cal L} /
a_B$, where the Coulomb barrier is effectively impenetrable
(opaque),~\cite{Andrews_76} $|t(q)| \ll 1$, due to the strong
backscattering at an exponentially short approach distance, $x \sim
q^{-1} \exp(-q a_B)$. We conclude that while the tails of the Coulomb
potential act as a small perturbation at momenta $q \sim k_F \equiv \pi
n$ and distances $x \sim a$, which are the most relevant for the
many-body problem, the strong {\it short-range\/} repulsion yields the
effective hard-core constraint for the charge dynamics. Therefore, in
the first approximation the Coulomb potential is equivalent to a very
thin and high barrier, i.e., to a $\delta$-function of a large strength.
Equation~(\ref{t_small_r}) supports this identification because up to
$O(1/{\cal L})$-terms, $t(q)$ coincides with the transmission amplitude
for the potential
\begin{equation}
U(x) = (\hbar^2 / m_\ast) c(k_F) \delta(x).
\label{U_contact}
\end{equation}
Note that in the opposite limit $r \ll 1$ of a low-energy scattering the
tails of the Coulomb potential cannot be ignored. The point of interest
$x \sim q^{-1}$ resides deeply inside a classically forbidden region of
the Coulomb barrier where the wavefunction $\phi$ depends exponentially
on $x$. This regime will not be important in what follows, but for
future reference we quote the counterpart of Eq.~(\ref{t_small_r}),
$|t(q)| \simeq ({\pi} / {\cal L}) \exp(-{\pi} / {q a_B})$.
Let us now apply the above ideas to the analysis of the many-body
problem.

Consider the limiting case $R = +0$ first. We have $c = \infty$ and
$t(q) = 0$ for all $q$; hence, the ground-state wavefunction $\Psi$ has
a node whenever coordinates $x_j$ of any particles coincide, $1 \leq j
\leq M$. Using an argument similar to Lieb-Mattis theorem,~\cite{Lieb_62}
one can show that $\Psi$ has no other nodes. Thus, it must have the form
\begin{eqnarray}
\Psi &=& \exp[W(x_1,\ldots,x_M)] \prod\limits_{Qi > Qj}
        \sin \frac{\pi}{L}(x_{Qi} - x_{Qj})\quad
\label{Psi_line_1}\\
 &\times& (-1)^Q \Phi(s_{1},\ldots, s_{M})
\label{Psi}
\end{eqnarray}
where $Q1$ through $Q{M}$ are the indices in the ordered list
of the electron coordinates $0 < x_{Q1} < \ldots < x_{Q{M}} < L$
(periodic boundary conditions are assumed), $(-1)^Q$ is the parity of
the corresponding permutation, $s_j$ is the spin of $Qj$\/th electron,
and $\Phi$ is the spin part of the wavefunction. The factor $\exp(W)$
incorporates the effect of weak $1/x$-tails of the interaction. Apart
from this, $\Psi$ coincides with the ground-state of electrons with
infinitely strong $\delta$-function repulsion,~\cite{Schlottmann_97}
i.e., the gas of impenetrable but otherwise free fermions (GIF). Below
we focus on correlation functions for which $W$ is not needed directly. 

Due to the strict impenetrability built into $\Psi$, particle exchanges
are forbidden, and so neither the parity factor $(-1)^Q$ nor spin are
dynamical degrees of freedom. (So, at $R = +0$ extra assumptions are
needed to fix $\Phi$). All correlation functions are slaved to those of
the density operator $\rho(x)$, e.g., the two-point cluster function,
\begin{equation}
h(x) = a^2 \langle \Psi | \rho(x) \rho(0)| \Psi \rangle - 1,
\label{h_def}
\end{equation}
In the GIF ($W = 0$) $h(x)$ is the same~\cite{Schlottmann_97} as in
the spin-polarized Fermi gas with the Fermi momentum $k_F = \pi n$,
\begin{eqnarray}
           h(x) &=& -\sin^2 (k_F x) / (k_F x)^2,
\label{h_GIF}\\
   \tilde{h}(q) &=& -a + \theta(2 k_F - q) q / (2 k_F n),
\label{h_GIF_q}
\end{eqnarray}
where $x, q^{-1} \ll L$, $\theta(z)$ is the step-function, and
henceforth the tilde denotes the Fourier transform.

The tails of the Coulomb potential cause a correction $\delta h(x)$ to
Eq.~(\ref{h_GIF}). From the analysis of the two-body problem, we expect
that at not too small $x$, $\delta h(x)$ admits a diagrammatic expansion
in $r_s$. Since the term $(-1)^Q \Phi$ does not affect the dynamics,
this expansion has identically the same form as for one-component
fermions, so that the standard calculation yields, in the leading order
%
\begin{eqnarray}
\delta\tilde{h} &\simeq& -\frac{r_s}{\pi^2}
\frac{q}{k_F n} \ln \frac{2 k_F}{q},\quad\: q \ll 2 k_F,
\label{delta_h_small_q}\\
&\sim& \delta{h}(2 k_F) + \frac{r_s |z|}{\pi^2 n} \ln^2 |z|,
\:\: z = \frac{2 k_F - q}{2 k_F} \to 0,\quad\quad
\label{delta_h_2k_F}\\
&\simeq& \frac{8}{\pi^2} \frac{r_s}{n}
\frac{k_F^4}{q^4},
\quad q \gg 2 k_F.
\label{delta_h_large_q}
\end{eqnarray}
As one can see, $\delta{h}(x)$ is a small correction to $h(x)$
[Eq.~(\ref{h_GIF})] up to an exponentially large distance $x_\ast \sim a
\exp({\pi^2} / {2 r_s})$. We show below that these first-order results
for $\delta h(x)$ smoothly match at $x \sim x_\ast$ with the asymptotic
long-distance behavior of $h(x)$ computed from the LL theory, which is
supposed to resum the perturbative series to all orders. We
conclude that at $r_s a \ll x \ll x_\ast$ the first-order
perturbation theory applies and that the charge correlations in the
CTG are indeed no different from those of free spinless fermions.

This result suffies to demonstrate that
the compressibility of the CTG is negative. We define the inverse
compressibility by $\varkappa^{-1} = {d^2 \varepsilon} / {d n^2}$, where
$\varepsilon(n)$ is the energy density of the system. To make it finite
the Hartree term (interaction with a neutralizing uniform backround) must
be subtracted away. This is an important difference from the case of
short-range interactions. To the leading order in $r_s$, $\varepsilon$
is equal to the kinetic energy of the spin-polarized Fermi gas plus the
potential energy evaluated using Eq.~(\ref{h_GIF_q}) for $\tilde{h}(q)$.
This simple calculation gives
%
\begin{equation}
\varkappa^{-1} = -\frac{2 e^2}{\kappa} \left[{\cal L}
 - \frac{\pi^2}{4 r_s} + \ln \left(\frac{2 r_s}{\pi}\right)
 - \gamma\right],
\label{kappa_CTG}
\end{equation}
which is indeed negative at $1 / {\cal L} \ll r_s \ll 1$. Here $\gamma$
is the Euler constant.~\cite{Gradshteyn_Ryzhik} With our definition of
$\varkappa$, its negative sign does not imply any instability of the CTG
towards, e.g., phase separation. The phase separation would cost a
large Hartree charging energy that would outweigh any gain due to
negative $\varkappa^{-1}$ term.~\cite{Comment_on_negative}

Let us briefly make a connection with the LL theory. The above result
for $\varkappa$ enters the LL machinery through the charge stiffness
parameter
\begin{equation}
K_\rho(q) = \sqrt{\frac{(\pi / 4) \hbar v}
                       {\tilde{U}(q) + \varkappa^{-1}}},
\quad v = \pi \frac{\hbar n}{m_\ast}.
\label{K_rho}
\end{equation}
Unlike the ``classical'' definition,~\cite{Haldane_81} in a Coulomb
LL~\cite{Schulz_93, Egger_98, Wang_01} $K_\rho$ depends not only on
$\varkappa$ but also on the interaction potential $\tilde{U}(q) \sim -2
(e^2 / \kappa) \ln R q$. This is again because the total energy cost of
the charge build-up is the sum of the negative $\varkappa$-term
(correlation energy~\cite{Comment_on_HF}) and the large postitive
Hartree term (electrostatic energy). $K_\rho$ shows up [through $v_c(q)
= v_F / K_\rho(q)$] in the dispersion $\omega = v_c(q) q$ of charge
mode. It also determines the low-$q$ behavior of the density correlation
function:
\begin{equation}
   \tilde{h}(q) + a \sim -{K_\rho(q)} q / ({k_F n}),\quad q \to 0.
\label{h_q_LLT}
\end{equation}
We notice that Eqs.~(\ref{delta_h_small_q}) and (\ref{h_q_LLT}) match at
$q \sim 1 / x_\ast$ and take it as evidence that Eq.~(\ref{h_q_LLT})
applies at $q < 1 / x_\ast$, while Eq.~(\ref{delta_h_small_q}) is valid
at $q > 1 / x_\ast$. Thus, the first-order perturbation theory is
sufficient for computing $h(x)$ at $x < x_\ast$. The LL theory can also
be used to calculate the crossover from Eq.~(\ref{delta_h_2k_F}) to a
different type of singularity in the immediate vicinity of $q = 2 k_F$.
If desired, one can use this to study in detail how the asymptotic
$\exp[-\sqrt{r_s \ln(x / x_\ast)}]$ decay~\cite{Schulz_93, Egger_98,
Wang_01} of the ``$2 k_F$'' ($2 a$-periodic) oscillations of $h(x)$ is
recovered at large $x$.

So far, we have discussed the case of an infinitely thin wire. If $R$ is
finite, $\Psi$ has nodes only at the coincident positions of the
same-spin particles. However, due to the opacity of the Coulomb barrier
in the CTG, $h(x)$ is perturbed very slightly. A much more important
difference is that particle exchanges become allowed and the spin
acquires some dynamics. Thus, it becomes meaningful and interesting to
determine $\Phi$. Since particle exchanges are still highly suppressed,
only those between nearest neighbors are relevant. In the CTG they are
determined by the two-body transmission amplitude $t(q)$. Recall that
the orbital part of the scattered wave depends on the total spin ${\bf
S} = {\bf S}_1 + {\bf S}_2$ of the two colliding particles,
\begin{equation}
   \phi = e^{i \xi / 2} \pm e^{-i \xi / 2}
 + [t(q) - 1] (e^{i |\xi| / 2} \pm e^{i |\xi| / 2}),
\label{phi_spin_dependence}
\end{equation}
where the upper (lower) sign is for the singlet (triplet). To the order
$O(t)$, exactly the same asymptotic scattered wave would result from the
exchange coupling $U_{\text{eff}} = (\textbf{S}_1 \textbf{S}_2 - 1/4)
U_\sigma(x)$ if at all $k$ such that $|t(k)| \ll 1$ we have
$\tilde{U}_{\sigma}(k) = 2 ({\hbar^2} / i {m_\ast}) t(k) k +
\text{const}$, as can be readily verified via Born approximation.
Although the form of the short-range potential $U_\sigma(x)$ in the real
space is not unique, this has no significance to the first order in the
small parameter $t$. Therefore, the spin-spin interaction is captured by
the Hamiltonian
\begin{equation}
  H_{\sigma} = \frac{\hbar^2}{m_\ast} \int \frac{d k}{2 \pi}
\left(\textbf{s}_k \textbf{s}_{-k} - \frac14 \rho_k \rho_{-k}
\right)
 \text{Im}\,t(k) k,
\label{H_sigma_Hubbard}
\end{equation}
where the integration is to be done up to the ultraviolet cutoff
$k_\text{max} \lesssim c(k_F)$ and ${\bf s}_k$, $\rho_k$ are the
harmonics of the spin and charge densities at wavevector $k$.
The physical idea expressed by Eq.~(\ref{H_sigma_Hubbard}) is very
similar to that behind the familiar antiferromagnetic coupling $(4 t^2 /
U) (\textbf{S}_i \textbf{S}_{i + 1} - n_i n_{i + 1} / 4)$ in a
large-$U$ 1D Hubbard model.~\cite{Ogata_90}

Since the spins are slow degrees of freedom, we can average $H_{\sigma}$
over the fast orbital motion to obtain the usual 1D $S = 1/2$ Heisenberg
model
\begin{equation}
{\textstyle H_\sigma = J \sum_j \textbf{S}_j \textbf{S}_{j + 1}},
\:\:\: J = \frac{\hbar v}{2 \pi^2}\int \tilde{h}(k)
 \text{Im}\,t(k) k d k.
\label{J}
\end{equation}
Therefore, the spin wavefunction $\Phi$ for the ground and excited
states are given by the appropriate Bethe ansatze.~\cite{Mattis_book}
Substituting Eq.~(\ref{t_small_r}) into Eq.~(\ref{J}) and keeping only
terms $O(1/{\cal L})$, we obtain 
\begin{equation}
J = \frac{\pi^2}{3} \frac{\hbar^2}{m_\ast}
 \frac{n^3 a_B}{{\cal L} + \ln {r_s}},
\quad  \frac{1}{\cal L} \ll r_s \ll 1.
\label{J_CTG}
\end{equation}
In the CTG, $J \ll E_F \equiv \hbar^2 k_F^2 / 8 m_\ast$, as expected. The
velocity of the spin excitations in 1D Heisenberg model
is~\cite{Mattis_book} $v_\sigma = (\pi / 2)(J a / \hbar)$ and the spin
susceptibility per unit length is $\chi_\sigma = (g \mu_B / 2)^2 (2 /
\pi \hbar v_\sigma)$. This implies that $\chi_\sigma$ of the CTG exceeds
that of the Fermi gas $\chi_\sigma^0$ (where $v_\sigma = v /
2$) by the large factor ${\chi_\sigma} / {\chi_\sigma^0} = {v} /
{2 v_\sigma} \approx r_s {\cal L} \gg 1$.

The low-temperature electron specific heat of a CTG is determined by the
velocities of the charge and the spin modes~\cite{Schlottmann_97}
and is dominated by the latter, $C_e = ({\pi} / {3}) (k_B T / \hbar
v_\sigma)
%
\sim (2 / \pi^2) (\kappa / e^2) (r_s^2 / {\cal L}) k_B T$.
%
This is large compared to $C_e$ in the Fermi gas because of the
smallness of $v_\sigma$. Obviously, $v_\sigma$ is relevant for many
observables. To verify that our strategy for finding $v_\sigma$ is
correct, we made sure that it reproduces the known exact
results~\cite{Ogata_90,Schlottmann_97} for the 1D Hubbard model and for
electrons with the contact interaction~(\ref{U_contact}),
\begin{equation}
v_\sigma \simeq (\pi^3 / 3) (\hbar n^2 / m_\ast c), \quad n \ll c.
\label{v_sigma_GIF}
\end{equation}
\noindent\textit{A gated wire\/}.--- Properties similar to those of the
CTG may also be exhibited by a modestly thin 1D wire, $R \lesssim a_B$,
if, instead of a high-$\kappa$ dielectric, it is positioned a small
distance $D$ away from a metallic gate. In that case one can model the
interaction potential by $U(x) = e^2 / |x| - e^2 / \sqrt{x^2 + 4 D^2}$
at $|x| \gg R$. This model was studied recently by H\"ausler~\textit{et
al.\/}~\cite{Hausler_02} who surmised that $K_\rho \to 1/2$ and
$v_\sigma \propto n^2 / \ln(D / R)$ at low $n$. In the regime $a \gg D^2
/ a_B \gg a_B$ the validity of these statements can be examined in a
controlled fashion. The interaction potential is short-range, $U(x)
\propto x^{-3}$ at $x \sim a$, and opaque, so the system is in the GIF
limit. After a straightforward calculation of $t(q)$ one finds that $c$
in Eq.~(\ref{t_small_r}) has to be replaced by~\cite{Matveev_indep}
\begin{equation}
c = A \frac{a_B}{D^2} \exp\left\{\sqrt{\frac{\pi D}{2 a_B}}
\left[\frac{\Gamma(1/8)}{\Gamma(5/8)} + \frac{\Gamma(5/8)}{\Gamma(9/8)}
\right]\right\},
\label{c_gated_wire}
\end{equation}
where coefficient $A \sim 1$ depends on the behavior of $U(x)$ at $x
\sim R$, and $\Gamma(z)$ is Euler's gamma
function.~\cite{Gradshteyn_Ryzhik} Since the interaction potential is
now short-range, subtraction of the Hartree term is no longer necessary.
We redefine $\varkappa^{-1} + \tilde{U}(q = 0) \to \varkappa^{-1}$ and
obtain $\varkappa^{-1} = \pi^2 e^2 a_B n$. Equation~(\ref{K_rho}) then
implies that $K_\rho \simeq 1/2$; however, the spin velocity, which can
be found by substituting Eq.~(\ref{c_gated_wire}) into
Eq.~(\ref{v_sigma_GIF}), differs from the surmise of
Ref.~\onlinecite{Hausler_02}.

\noindent\textit{Experimental manifestations\/}.--- The predicted large
difference of spin and charge mode velocities ($v_c$ and $v_\sigma$) can
be verified by momentum-resolved photoemission~\cite{Claessen_02} or
tunneling.~\cite{Auslaender_02} Since CN is currently the best candidate
for realizing the Coulomb Tonks Gas regime, we have to mention here that
the bands of a pristine CN have a two-fold valley degeneracy. This can
be accommodated into the model via effective spin-$1/2$ operators
$\textbf{T}_i$. Each two-body term in Eq.~(\ref{J}) should then be
replaced by $J (\textbf{S}_i \textbf{S}_{i + 1} + 1 / 4) (2 \textbf{T}_i
\textbf{T}_{i + 1} + 1 / 2)$. The resultant $H_\sigma$ remains
integrable~\cite{Li_99} and one finds that now $v_\sigma = (\pi^3 / 6)
(\hbar n^2 / m_\ast c)$. Transport is another powerful probe of the CTG
and Wigner crystal regimes. In the temperature window $J \ll k_B T \ll
E_F$, with $J \sim 1\,{\rm K}$ (a crude estimate), ballistic conductance
$G$ through a CN should be unusual. Adopting the theory of
Ref.~\onlinecite{Matveev_04} to the two-valley case, we find $G = e^2 /
h$, which is $1/4$ of the non-interacting result, i.e., the
``0.25-anomaly.'' On the other hand, $G$ measured through a CN with
resistive (tunneling) contacts may display an unusual power-law
decrease~\cite{Cheianov_04} with $T$. The negative compressibility would
reduce the charging energy of finite wires, and so Eq.~(\ref{kappa_CTG})
for $\varkappa$ can be tested by careful Coulomb blockade experiments on
the CN quantum dots. Finally, the strong electron correlations
can also be observed~\cite{Scanning} in imaging microscopy: instead of
the $4 a$-periodic Friedel oscillations of a Fermi gas, charge density
near boundaries of defects would oscillate with period $a$, i.e., four
times smaller, in the CTG. With further descrease of $n$, these
oscillations would become strongly anharmonic as appropriate for a pinned
Wigner crystal. Detailed quantitative predictions for all such
measurements will be discussed elsewhere.

This work is supported by the C. \& W.\ Hellman Fund and the A. P. Sloan
Foundation. I am indebted to L.~S.~Levitov for many valuable insights
into the problem and to D.~Arovas, G.~Fiete, W.~H\"ausler, and
E.~Pivovarov for discussions and comments.



\begin{thebibliography}{99}
\bibitem{Matveev_indep} Equation~(\ref{c_gated_wire}) was
independently derived in Ref.~\onlinecite{Matveev_04}(b).
\end{thebibliography}
\end{document}